# "Vertically coupled polymer microracetrack resonators for Label-Free Biochemical sensors"

We report on the efficient design and fabrication of polymeric microracetrack optical resonators for label-free biosensing purposes. Vertically-coupled microresonators immersed in deionised water display high Q-factors (>35000)and finesses up to 25. A surface sensing experiment performed with these microresonators using 5-TAMRA cadaverine as a test molecule demonstrated both the high sensitivity and low detection limit of our device.

# Vertically Coupled Polymer Microracetrack Resonators for Label-Free Biochemical Sensors

C. Delezoide, M. Salsac, J. Lautru, H. Leh, C. Nogues, J. Zyss, M. Buckle, I. Ledoux-Rak
and C.T. Nguyen, Member, IEEE

*Abstract*— We report on the efficient design and fabrication of polymeric microracetrack optical resonators for label-free biosensing purposes. Vertically-coupled microresonators immersed in deionised water display high Q-factors (> 35000) and finesses up to 25. A surface sensing experiment performed with these microresonators using 5-TAMRA cadaverine as a test molecule demonstrated both the high sensitivity and low detection limit of our device.

## I. INTRODUCTION

Label-free biochemical sensors integrating optical microresonators represent an emerging technology that has been under intensive investigation lately. The detection mechanism of these sensors relies on the sensitivity of the optical mode guided inside the core of the microresonator to its surroundings via the evanescent field, magnified by the long effective interaction length obtained near resonance. Thus, it can be used either to measure very small concentrations of single chemical species in solution, using the sensitivity to the bulk refractive index (RI) of the solution acting as a cladding material, and/or, to study targeted biomolecular interactions via surface adsorption [1]. In the latter scheme, functionalization of the free surface of the resonator facilitates the obtention of both high sensitivity and high specificity in sensing. Two techniques can be used to interpret modifications of the resonator environment, both based on the shift of resonances accompanying effective index variations. This shift can be measured either using spectral scanning of the transmitted intensity through the bus waveguide coupled to the resonator or by recording the variations of the same transmitted intensity at a fixed wavelength. Accordingly, biosensors can be characterized by their sensitivities $S_{\lambda,I}$, defined as the ratio between either the variation of the resonant wavelength or the variation of intensity at a given wavelength, per molar or mass quantity of the analyte. An equally important parameter is their detection limit $DL_{\lambda,I} = R_{\lambda,I}/S_{\lambda,I}$. $R_{\lambda,I}$, the resolution of the sensor, characterizes the smallest resonant wavelength or intensity variations that can be accurately measured. This term takes into account both spectral resolution of the source and signal-to-noise ratio (SNR) of the sensor. In biosensing, detection limit (DL) describes the smallest amount of analyte that the sensor can accurately quantify [2]. It was shown that a higher Q-factor of the resonator meant more sensitivity for the sensor [3]. The sensitivity also depends on the spatial extension of the evanescent wave, defining the interaction strength between light and the analyte. Therefore, it is important to optimize these parameters in order to achieve high sensitivity, a task that is mainly application-dependent.

Beyond the sensitivity and DL issues, our designs were orientated towards the production of low-cost, easy to fabricate, easy to use and robust devices to be competitive with already existing and relatively common platforms such as SPR based sensors.

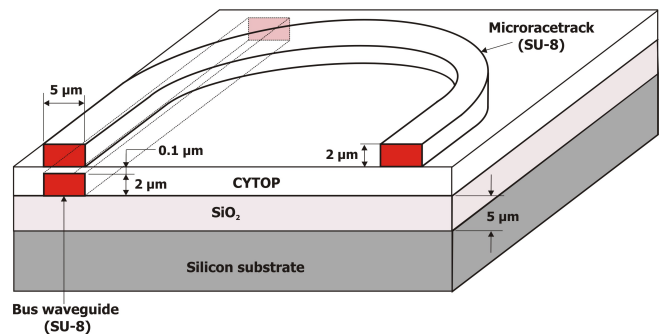

Fig. 1. Vertically-coupled polymeric microracetrack resonator

## II. DESIGN AND FABRICATION

To develop optical label-free biosensors, polymer materials offer several advantages. Amongst these are the low-cost and ease of fabrication. Additionally, the relatively low RIs of polymer materials results, in comparison with SOI technologies, in lower scattering loss on sidewalls, increased SNR and a higher coupling efficiency to the bus waveguide. Moreover, for surface detection applications, polymer surfaces can be easily modified to immobilize a wide range of biomolecules. Finally, in order to introduce and control the analytes sensed by the resonator, biosensors usually integrate

Manuscript received July 18, 2011. This work was supported in part by the CNano Ile-de-France under Grant n°13 Bioptofluidic.
C. Delezoide, I. Ledoux-Rak and C. T. Nguyen are with LPQM, UMR 8537 CNRS, ENS Cachan, 94235 Cachan, France (Phone: 33-1-47405557; fax: 33-1-47405567; e-mail: ctnguyen@ lpqm.ens-cachan.fr).
M. Salsac, H. Leh, C. Nogues and M. Buckle are with LBPA, UMR 8113 CNRS, ENS Cachan, 94235 Cachan, France (e-mail: msalsac@lbpa.ens-cachan.fr).
J. Lautru and J. Zyss are with Institut d'Alembert, ENS Cachan, 94235 Cachan, France (e-mail: zyssr@lpqm.ens-cachan.fr).

microfluidic circuitry, forming optofluidic systems. Polymers are well compatible with this technology, facilitating the upgrade of photonic circuits into optofluidic cells.

A vertical coupling configuration, in which resonators are vertically coupled to the buried bus waveguide, as shown in Fig. 1, was chosen. The bus waveguide acts as an input/output channel while the racetrack works as the sensitive component. The vertical coupling configuration offers several advantages over lateral coupling. Firstly, the intensity coupling can be controlled to a finer degree as the vertical gap which separates the buried bus waveguide from the racetrack is easily obtained by controlled layer deposition and etching, and the coupling strength is insensitive to small misalignments of waveguides [4]. Secondly, in the lateral coupling configuration, sufficient coupling strength from and to the microresonator can only be achieved through high-resolution fabrication techniques such as e-beam and nanoimprint lithographies. In addition to being more expensive than the techniques used to implement the vertical coupling configuration, they also often require further fabrication steps to reduce surface roughness and obtain a comparable quality of resonators [5]. Also, the vertical coupling configuration, in the context of biochemical sensing, is advantageous since the coupling region is completely isolated from the analyte [6]. This drastically reduces the influence of local refractive indices modifications on the shape of the resonance, which makes it possible to study a variety of interactions, with different buffer solutions, without loss of sensitivity. Finally, the vertical separation of the bus waveguide to the microracetrack resonator facilitates the integration of microfluidic channel roof upon the lower cladding of optical microresonators.

Optimization of the geometries of the photonic elements was performed using a combination of 3D-BPM, 2/3D-FDTD and ADI method for mode solving, all with the Optiwave software suite. Typical geometry of the fabricated devices is shown in Fig. 1. A thick layer of silica as substrate for the bus waveguides reduces the substrate loss. The core material, for both bus and resonator waveguides, is the negative epoxy photoresist SU-8, offering easy fabrication and high aspect ratio. A commercial amorphous fluoropolymer, CYTOP, was used as upper cladding of the straight bus waveguide and also lower cladding of the racetrack resonator. The RIs of silica, SU-8 and CYTOP, measured at the 1.55 µm wavelength by ellipsometry, are respectively 1.442, 1.564 and 1.333. Typical width and height of the waveguides are w = 5 µm and h = 2 µm respectively. Different curvature radii R and straight portions lengths $L_c$ were tested, between 80 and 220 µm for R, and between 0 and 220 µm for $L_c$. A typical value for the gap is around 100 nm.

Devices were fabricated by adopting a standard procedure combining near UV lithographies for SU-8 waveguides and Reactive Ion Etching (RIE) to control the CYTOP gap thickness. A 2 µm-thick layer of SU-8 was spin-coated on the Si/SiO$_2$ substrate. Desired patterns were obtained by UV exposure through a first lithography mask and developed to obtain the bus waveguides. Subsequently, a layer of CYTOP was deposited, also by spin-coating, and the desired thickness of the gap was obtained by time-controlled oxygen plasma RIE. After this step, a second 2 µm-thick SU-8 layer was spin-coated over the etched CYTOP layer and a second UV lithography was performed to obtain the microresonators. Finally, all devices were cured at 80°C for 4 hours.

This fabrication process, in addition to being simpler, hence less expensive, and faster that previously reported [6], has the decisive advantage to minimize the roughness of the surface of the resonators, which usually causes a large part of the in-cavity optical losses.

### III. CHARACTERIZATION

Fabricated devices were characterized using a Tunics-Plus tunable laser source emitting in the 1500-1640 nm range, with a ±40 pm accuracy. Polarization of the input beam was controlled by a combination of a broadband polarizer and a half-wave plate. Coupling of the input beam and collection of the output beam were respectively carried out with x40 and x20 microscope objectives, both treated to minimize the parasitic reflections in the IR. Temperature of the device was controlled via a thermostatic mount integrating a thermal sensor and thermo-electric cooler devices connected to a PID servo system.

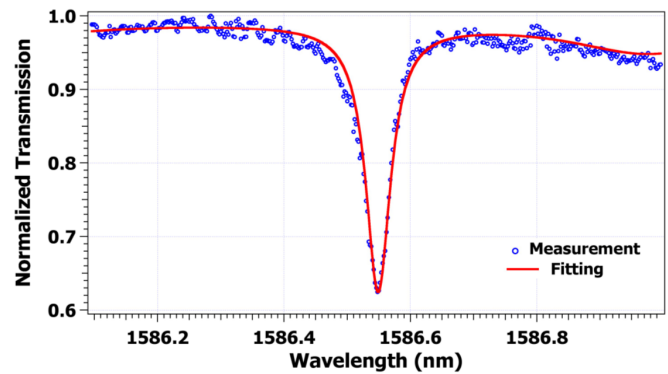

Fig. 2. Detailed measurement of a resonance peak from the transmission spectrum (TE polarization) of the polymeric microracetrack resonator

The different geometries of racetrack microresonators were characterized at 22°C using deionised (DI) water as cladding. The characterization bench exhibited transmission coefficients reaching 4% at maximum from the source to the power-meter. The best result, shown in Fig.2, was obtained from a microracetrack resonator with $L_c$ = 120 µm and R = 180 µm. The device displayed, around 1584 nm, resonance peaks with typical FWHM of 45 pm and a FSR of 1.15 nm, which corresponds to a Q-factor of 35,200 and a finesse equal to 25. This Q-factor, measured with DI water as cladding, is, to the best of our knowledge, the highest reported for a polymeric micro-ring/racetrack resonator in both vertical [6] and lateral [5], [7] configurations. Repeating the same fabrication process, similar results were obtained for this geometry in subsequent products. Increasing only the vertical gap separation produced resonators with lower Q-factors, between 10,000 and 20,000, but with better extinction ratios, up to 12 dB.

## IV. Demonstration of Surface Binding Sensing

Surface binding detection experiments were carried out using solutions with various concentrations of Tetramethylrhodamine-5-carboxamide (5-TAMRA) cadaverine. This molecule contains a fluorophore attached to an amine group which allows quantification via fluorescence analysis of the amount of material attached to a surface. To enable binding of the 5-TAMRA cadaverine on a prepared SU-8 surface, a novel approach, based on UV-ozone surface treatment, was developed. The increasing quantity of carboxyl groups produced by longer treatments, measured with IR-spectroscopy as shown in Fig.3, illustrates the higher capacity of the treated surface to bind 5-TAMRA cadaverine molecules through peptide bonds. A study of the fluorescence signal produced by spots of 5-TAMRA cadaverine bound to treated surfaces of SU-8 also demonstrated that a short exposure to UV/ozone was sufficient to obtain maximal reactivity, as longer treatments created more carboxyl groups in the bulk of the material but not on its surface.

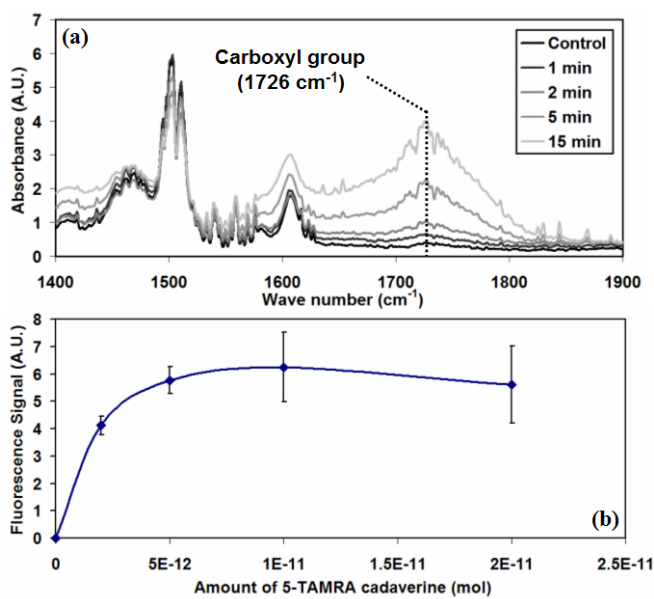

Fig. 3. Effect of the UV/ozone treatment on SU-8 measured by FTIR spectrometry (a) and calibration curve for the quantification of 5-TAMRA cadaverine on SU-8 surface (b)

The effect of the surface densities of immobilized 5-TAMRA cadaverine on the spectral response of the resonators was studied under the same conditions used for the optical characterization previously described. Fluorescence allowed an approximate quantification of the surface density at each measurement point with a 30 % error margin. Results are displayed in Fig. 4. In this experiment, a 15 pm shift corresponded to a density of 53,900 molecules/mm$^2$ on the sensing surface $\Sigma_{tot}$ of the resonator. With $\Sigma_{tot}$ = 0.0143 mm$^2$ and a molar mass M = 517 g/mol for the 5-TAMRA cadaverine, we can estimate the sensitivity obtained at the lower concentrations to $S_\lambda$ = 23 nm/fg. Considering the wavelength-stability of the laser source and the noise level, the resolution $R_\lambda$ of the sensor for these measurements was evaluated at 5 pm. Accordingly, the detection limit $DL_\lambda$ for this experiment was evaluated at approximately 0.22 ag which corresponds to 250 molecules bound on the surface of the resonator.

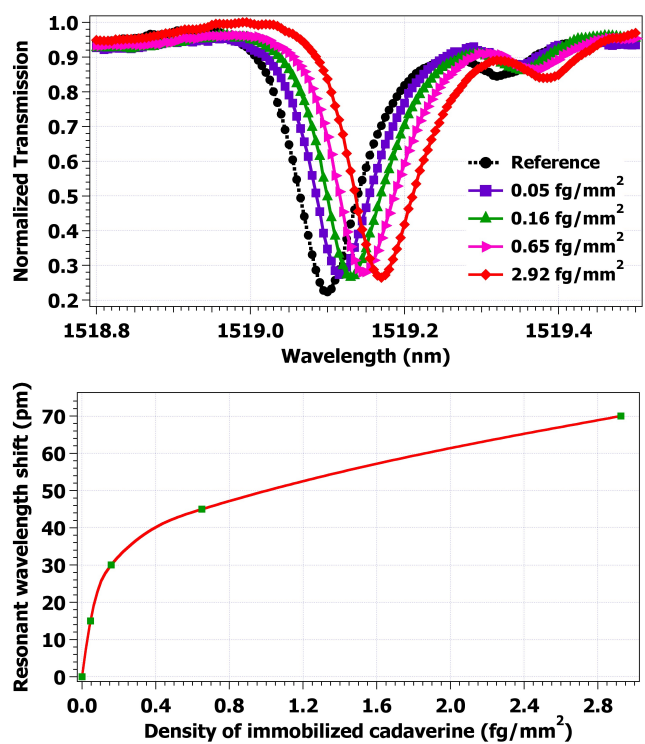

Fig. 4. Demonstration of surface sensing of binded 5-TAMRA cadaverine molecules. (a) Spectral responses for various mass surface densities; and (b), resonant wavelength shift as a function of the mass surface density

## V. Conclusion

An efficient design for a low-cost label-free biosensor was presented, and high sensitivity and low detection limit of the fabricated devices were demonstrated. Further improvement of our setup will be carried out with the addition of microfluidics, the reduction of the time response and surface engineering to achieve washable optofluidic cells able to study biological interactions with high specificity.